\documentclass[12pt]{article}
\usepackage{graphicx}

\newenvironment{enumerate*}%
  {\begin{enumerate}%
    \setlength{\itemsep}{1pt}%
    \setlength{\parskip}{1pt}}%
  {\end{enumerate}}

\newcommand\pubnumber{WSU--HEP--XXYY}
\newcommand\pubdate{\today}

\def\binp{Budker Institute of Nuclear Physics, 11, Lavrentieva str., Novosibirsk, 630090, Russia\\
Novosibirsk State University, 2, Pirogova str., Novosibirsk, 630090, Russia}
\def\support{\footnote{Speaker, on behalf of the Belle collaboration.}}

\def\Title#1{\begin{center} {\Large #1 } \end{center}}
\def\Author#1{\begin{center}{ \sc #1} \end{center}}
\def\Address#1{\begin{center}{ \it #1} \end{center}}

\newcommand\pubblock{\rightline{\begin{tabular}{l} \pubnumber\\
         \pubdate  \end{tabular}}}
\newenvironment{Abstract}{\begin{quotation}  }{\end{quotation}}
\newenvironment{Presented}{\begin{quotation} \begin{center} 
             PRESENTED AT\end{center}\bigskip 
      \begin{center}\begin{large}}{\end{large}\end{center} \end{quotation}}

\def\beq{\begin{equation}}
\def\eeq#1{\label{#1}\end{equation}}
\def\eeqn{\end{equation}}

\def\beqa{\begin{eqnarray}}
\def\eeqa#1{\label{#1}\end{eqnarray}}
\def\eeqan{\end{eqnarray}}

\let\bar=\overbar

\def\Dslash{\not{\hbox{\kern-4pt $D$}}}
\def\dslash{\not{\hbox{\kern-2pt $\del$}}}

\def\msb{{\bar{\ssstyle M \kern -1pt S}}}

\begin{document}
\begin{titlepage}
\pubblock

\vfill
\Title{Search for $B$ decays to final states with the $\eta_c$ meson}
\vfill
\Author{Anna Vinokurova\support}
% put in address(es) defined above
\Address{\binp}
\vfill
\begin{Abstract}
We report a search for $B$ decays to selected final states with the $\eta_c$ meson: 
$B^{\pm}\to K^{\pm}\eta_c\pi^+\pi^-$, $B^{\pm}\to K^{\pm}\eta_c\omega$, 
$B^{\pm}\to K^{\pm}\eta_c\eta$ and $B^{\pm}\to K^{\pm}\eta_c\pi^0$. The analysis 
is based on $772\times 10^6$ $B\bar{B}$ pairs collected at the 
$\Upsilon(4S)$ resonance with the Belle detector at the KEKB 
asymmetric-energy $e^+e^-$ collider.
We set 90\% confidence level upper limits on the branching fractions of the studied $B$ decay modes, 
independent of intermediate resonances, in the range $(0.6-5.3)\times 10^{-4}$. 
We also search for molecular-state candidates in the 
$D^0\bar{D}^{*0}-\bar{D}^0D^{*0}$, $D^0\bar{D}^0+\bar{D}^0D^0$ and 
$D^{*0}\bar{D}^{*0}+\bar{D}^{*0}D^{*0}$ combinations, neutral partners of the 
$Z(3900)^{\pm}$ and $Z(4020)^{\pm}$, and a poorly understood state $X(3915)$ 
as possible intermediate states in the decay chain, and set 90\% confidence level 
upper limits on the product of branching fractions to the mentioned intermediate states and decay 
branching fractions of these states in the range $(0.6-6.9)\times 10^{-5}$.
\end{Abstract}
\vfill
\begin{Presented}
The 7th International Workshop on Charm Physics (CHARM 2015)\\
Detroit, MI, 18-22 May, 2015
\end{Presented}
\vfill
\end{titlepage}
\def\thefootnote{\fnsymbol{footnote}}
\setcounter{footnote}{0}
%

%%%%%%%%%%%%%%%%%%%%%%%%%%%%%%%%%%
\section{Introduction}

The state $X(3872)$ was first observed by Belle in exclusive $B^+\to K^+\pi^+\pi^-J/\psi$ decays~\cite{Belle1}. The $X(3872)$ mass is close to the $m_{D^0}+m_{\bar{D}^{*0}}$ threshold, which engendered a hypothesis that this state may be a $D^0\bar{D}^{*0}$ molecule~\cite{Molecule}. Angular analysis of the $X(3872)\to J/\psi\pi^+\pi^-$ decay by LHCb~\cite{LHCb2} determined all its quantum numbers: $J^{PC}=1^{++}$. If $X(3872)$ is indeed a $D^0\bar{D}^{*0}$ molecule, there can exist other ``$X(3872)$-like'' molecular states with different quantum numbers. Some may reveal themselves in the decays to final states containing the $\eta_c$ meson. For example, a $D^0\bar{D}^{*0}-\bar{D}^0D^{*0}$ combination (denoted hereinafter by $X_1(3872)$) with quantum numbers $J^{PC}=1^{+-}$ would have a mass around $3.872$ GeV/$c^2$ and would decay to $\eta_c\rho$ and $\eta_c\omega$. Combinations of $D^0\bar{D}^0+\bar{D}^0D^0$, denoted by $X(3730)$, and $D^{*0}\bar{D}^{*0}+\bar{D}^{*0}D^{*0}$, denoted by $X(4014)$, with quantum numbers $J^{PC}=0^{++}$ would decay to $\eta_c\eta$ and $\eta_c\pi^0$. The mass of the $X(3730)$ state would be around $2m_{D^0}=3.730$ GeV/$c^2$ while that of the $X(4014)$ state would be near $2m_{D^{*0}}=4.014$ GeV/$c^2$.

Recently, a new charged state $Z(3900)^{\pm}$ was found in $Y(4260)$ decays by Belle~\cite{zbelle} and BESIII~\cite{zbes1}. Since this particle is observed in the decay to $\pi^{\pm}J/\psi$, it should contain at least four quarks. The $Z(3900)^{\pm}$ was confirmed in the decay to $\pi^{\pm}J/\psi$ by an analysis of CLEO-c data~\cite{zcleo} that also reported evidence for its neutral isotopic partner $Z(3900)^0$. Another exotic charged state $Z(4020)^{\pm}$ was observed by BESIII in decays to $\pi^{\pm}h_c$~\cite{zbes3} and $(D^*\bar{D}^*)^{\pm}$~\cite{zbes4}. There are some indications from these analyses that the spin and parity of the charged states might be $J^P=1^+$.

The near-threshold enhancement in the $\omega J/\psi$ invariant mass distribution named $Y(3940)$ was first observed by Belle in exclusive $B\to K\omega J/\psi$ decays~\cite{x39151}. Later, in the same decay mode, BaBar discovered $X(3915)$~\cite{x39152}. The parameters of $Y(3940)$ are consistent with those of $X(3915)$, so they are considered to be the same particle. The quantum numbers of $X(3915)$ are claimed to be $J^{PC}=0^{++}$, but its nature is still undetermined, and there are several interpretations describing this state~\cite{x39156}. 

To search for the particles described above, we reconstruct $\eta_c$ mesons via the $K_S^0K^{\pm}\pi^{\mp}$ mode and study the following four decays of charged $B$ mesons:
\begin{enumerate*}
\item the ($\pi^+\pi^-$) decay mode: $B^{\pm}\to K^{\pm}X\to K^{\pm}(\eta_c\pi^+\pi^-)$,
where we look for $X_1(3872)$, $Z(3900)^0$ and $Z(4020)^0$;
\item the ($\omega$) decay mode: $B^{\pm}\to K^{\pm}X\to K^{\pm}(\eta_c\omega)$,
where we look for $X_1(3872)$;
\item the ($\eta$) decay mode: $B^{\pm}\to K^{\pm}X\to K^{\pm}(\eta_c\eta)$,
where we look for $X(3730)$, $X(4014)$ and $X(3915)$;
\item the ($\pi^0$) decay mode: $B^{\pm}\to K^{\pm}X\to K^{\pm}(\eta_c\pi^0)$,
where we look for $X(3730)$, $X(4014)$ and $X(3915)$.
\end{enumerate*}

The analysis is based on a data sample that contains $772\times 10^6$
$B\bar{B}$ pairs, collected with the Belle detector at the KEKB 
asymmetric-energy $e^+e^-$ collider.

\section{$B^{\pm}\to K^{\pm}\eta_c+{\rm hadrons}$}

\begin{figure}[htb]
\centering
\includegraphics[height=1.8in]{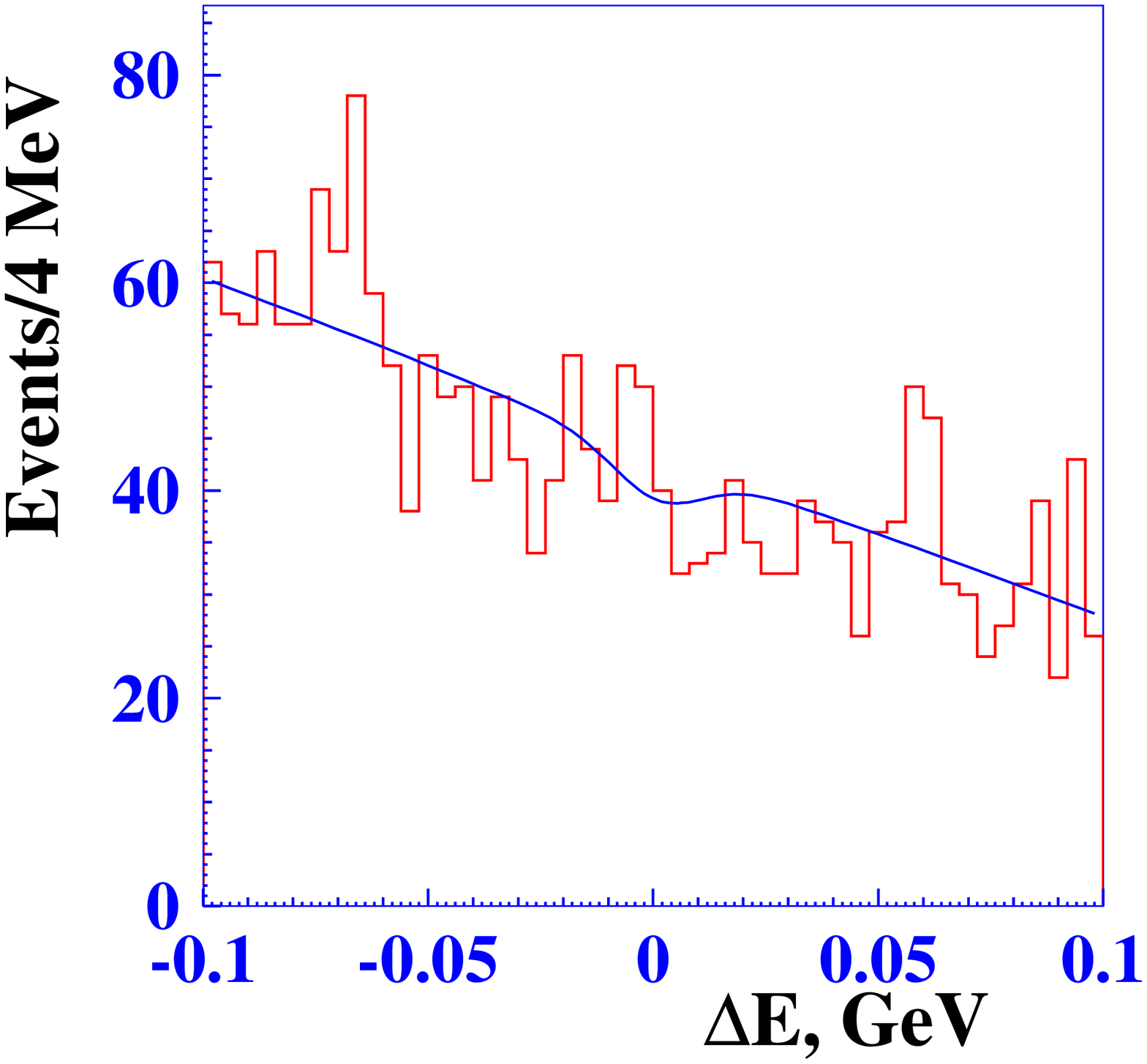}
\includegraphics[height=1.8in]{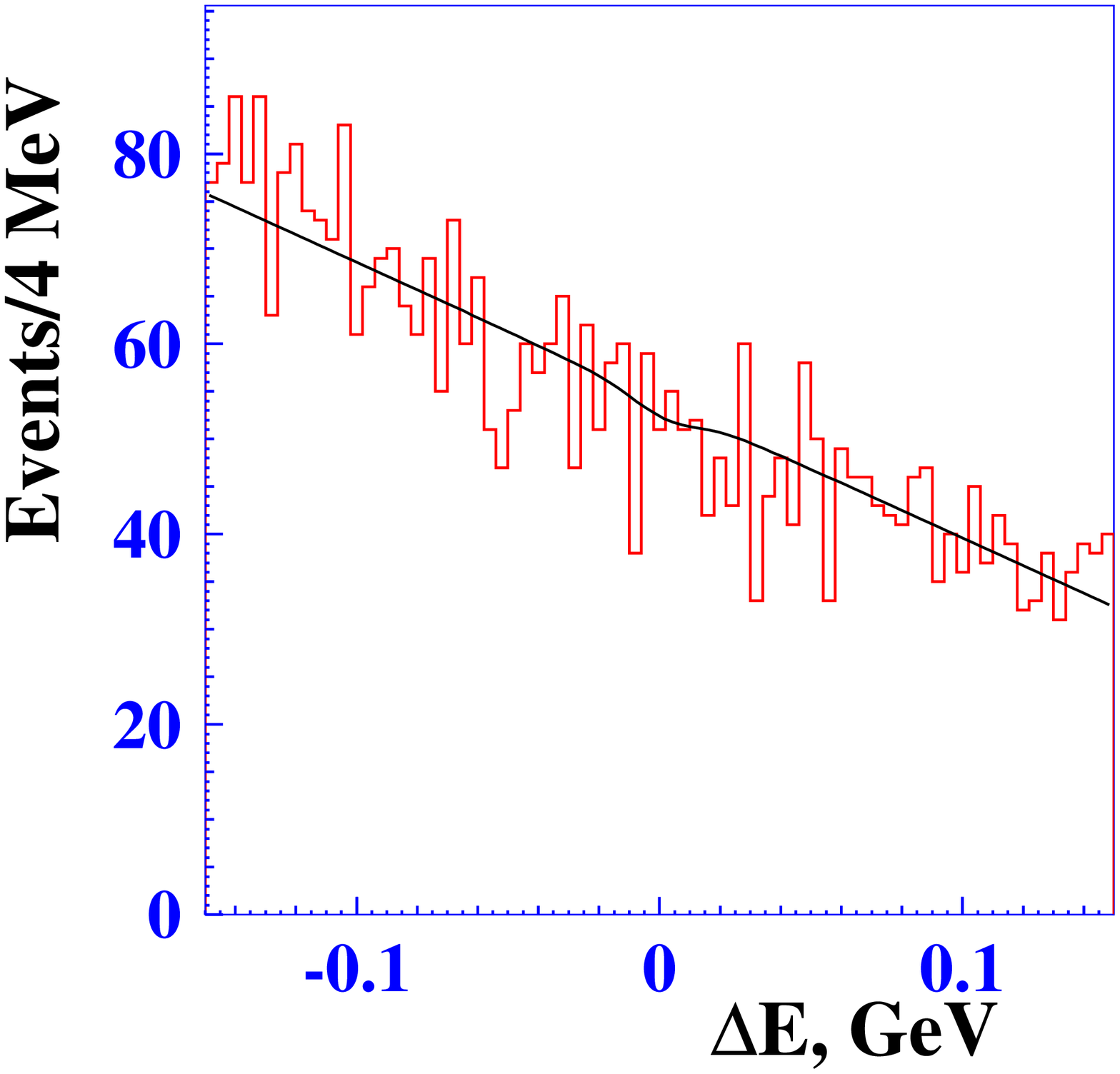}
\includegraphics[height=1.8in,origin=c,angle=0]{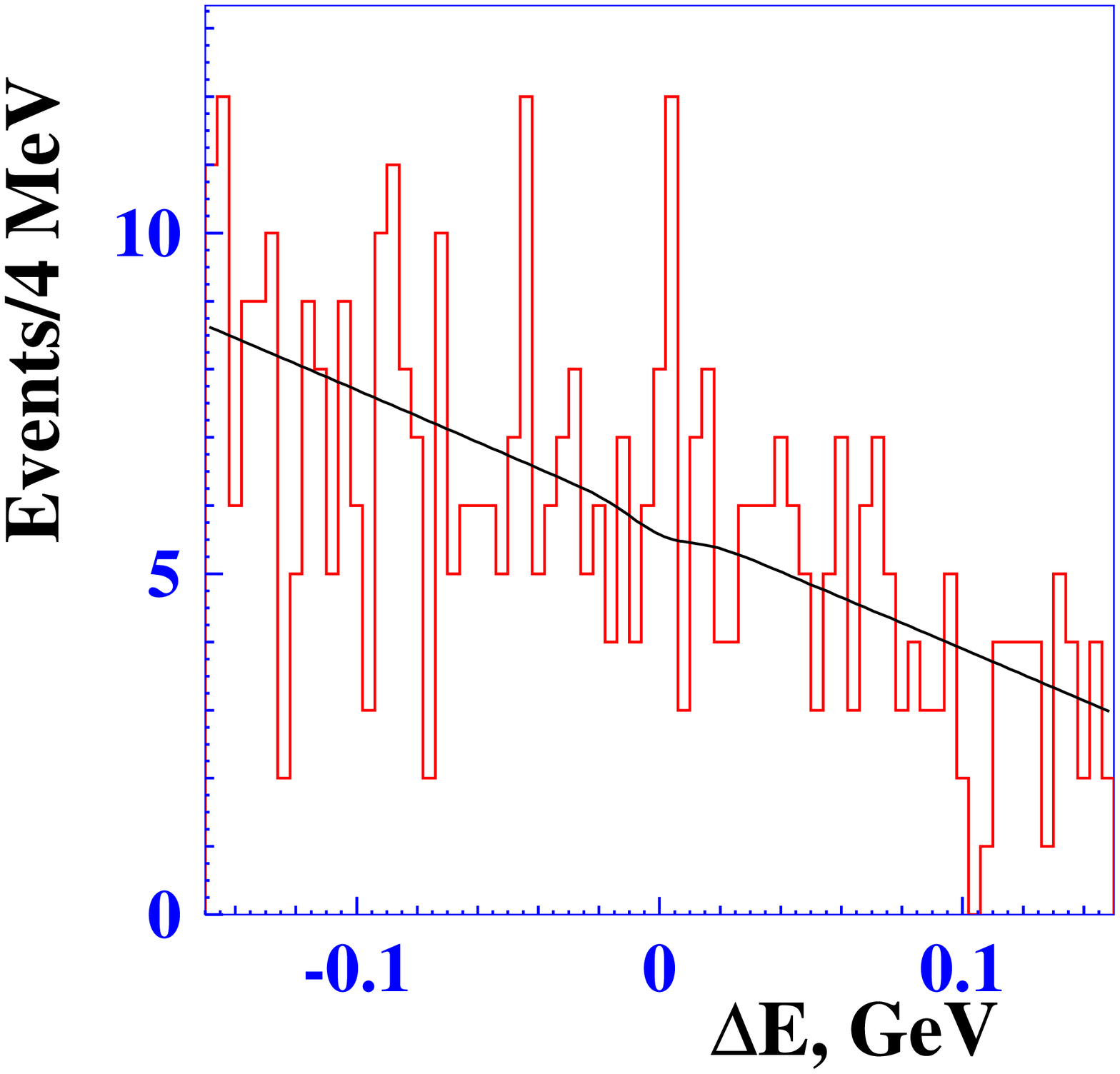}
\caption{The $\Delta E$ distribution for the decay $B^{\pm}\to K^{\pm}\eta_c\omega$ (left). The combined fit projections of the $\Delta E$ distributions in case of the $\eta\to\gamma\gamma$ (middle) and $\eta\to\pi^+\pi^-\pi^0$ (right) modes for the decay $B^{\pm}\to K^{\pm}\eta_c\eta$.}
\label{pic:23}
\end{figure}

\begin{figure}[htb]
\centering
\includegraphics[height=1.8in]{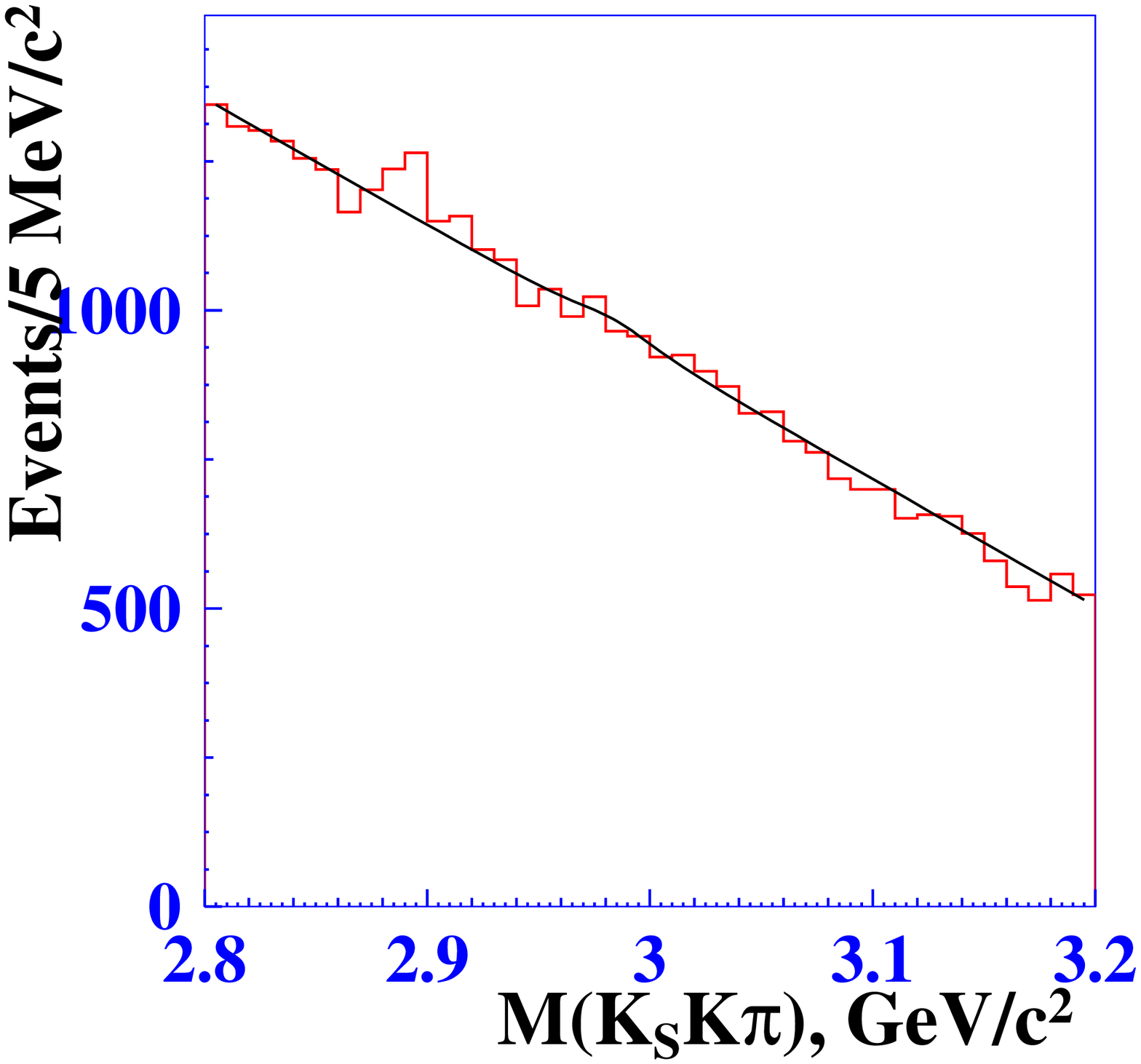}
%\hfill
\includegraphics[height=1.8in,origin=c,angle=0]{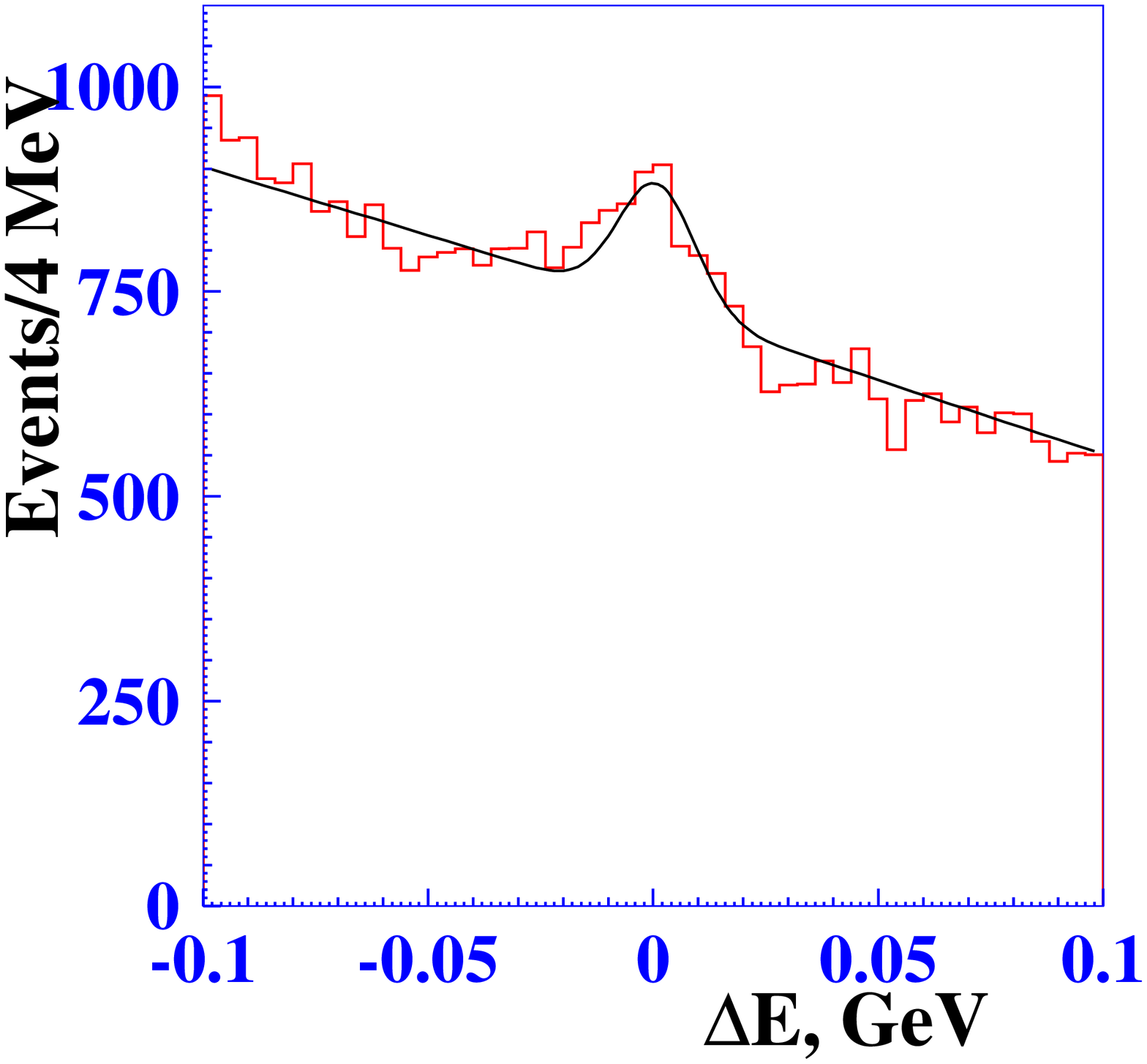}
\vfill
\includegraphics[height=1.8in]{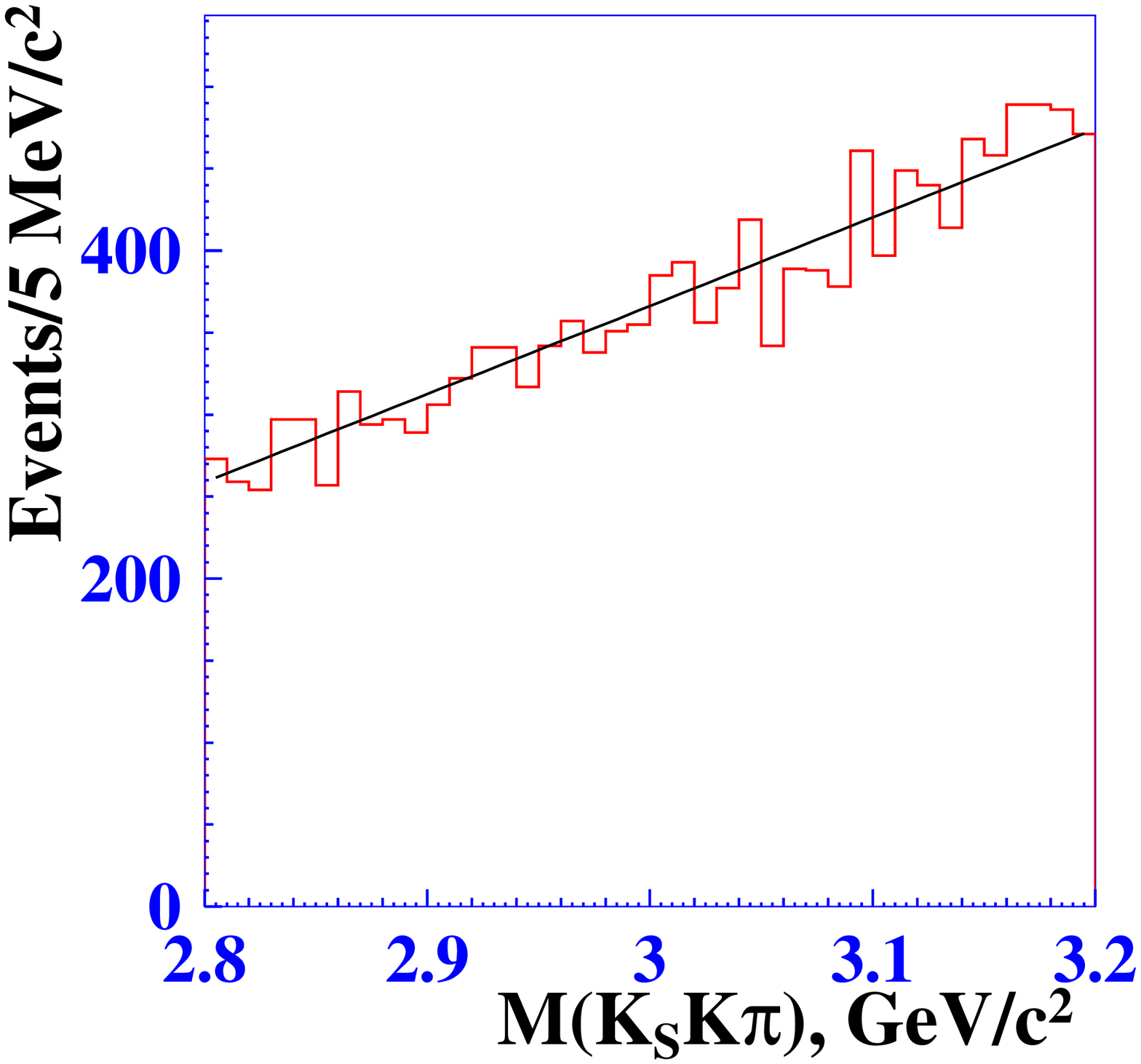}
\includegraphics[height=1.8in,origin=c,angle=0]{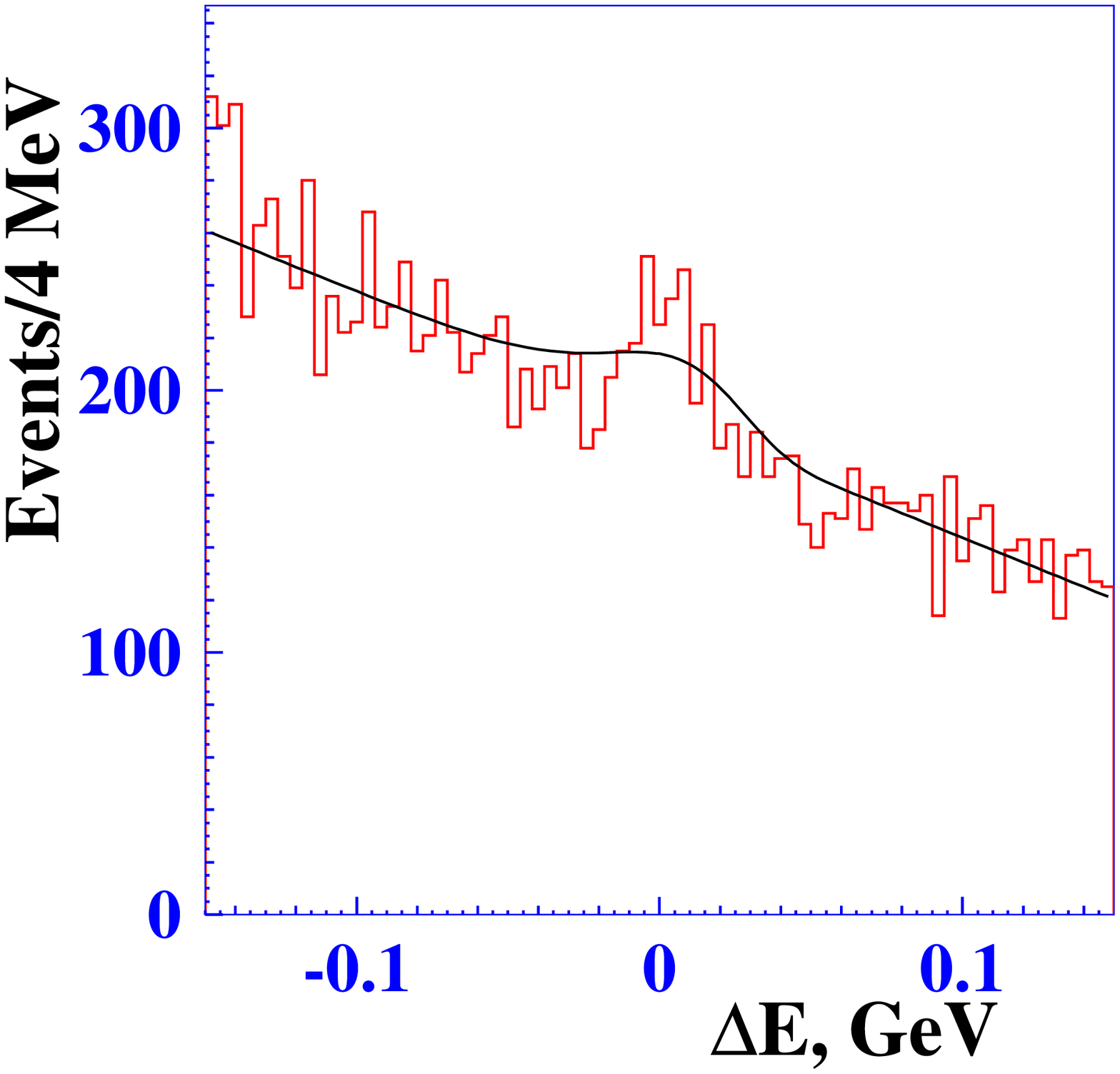}
\caption{Projections of the two-dimensional fit in $K_SK\pi$ invariant mass (left) and $\Delta E$ (right) for the decays $B^{\pm}\to K^{\pm}\eta_c\pi^+\pi^-$ (top) and $B^{\pm}\to K^{\pm}\eta_c\pi^0$ (bottom). Each projection is plotted for events in the whole fitting range of the other projection.}
\label{pic:22}
\end{figure}

We search for ${D}^{(*)0}D^{(*)0}$ molecular-state candidates $Z(3900)^0$, $Z(4020)^0$, and $X(3915)$ in the following $B$ meson decays: $B^{\pm}\to K^{\pm}\eta_c\pi^+\pi^-$, $B^{\pm}\to K^{\pm}\eta_c\omega$, $B^{\pm}\to K^{\pm}\eta_c\eta$, and $B^{\pm}\to K^{\pm}\eta_c\pi^0$. 
To determine the branching fractions, we perform a binned maximum-likelihood fit of the $\Delta E$ distribution that is modelled by a peaking signal and featureless background.
For the ($\pi^+\pi^-$), ($\omega$) and ($\pi^0$) decay modes, the signal function is the sum of two Gaussians ($G$) 
and the background function is a linear polynomial. Here and in the following the detector resolution is taken into account. The $\Delta E$ distribution for the ($\omega$) mode is shown in Figure~\ref{pic:23} (left). In the ($\pi^+\pi^-$) and ($\pi^0$) modes, we observe some significant signal and so perform a two-dimensional fit of the $K_SK\pi$ invariant mass and $\Delta E$ distributions shown in Figure~\ref{pic:22}. Since the $\eta$ candidate is reconstructed in two decay modes, we perform a combined fit of the $\Delta E$ distribution corresponding to $\eta\to\gamma\gamma$ and $\eta\to\pi^+\pi^-\pi^0$ shown in Figure~\ref{pic:23} (middle and right). The fit results are summarized in Table~\ref{tab:21}.
\begin{table}[htb]
\begin{center}
\begin{tabular}{l|cc} 
Decay mode & Efficiency, \% & Yield \\
\hline 
$B^{\pm}\to K^{\pm}\eta_c\omega$ & $0.53\pm 0.01$ & $-41\pm 27$\\
$B^{\pm}\to K^{\pm}\eta_c\pi^+\pi^-$ & $2.84\pm 0.02$ &
$155\pm 72$ \\
$B^{\pm}\to K^{\pm}\eta_c\pi^0$ & $3.69\pm 0.01$ & $-1.9\pm 12.1$ \\
$B^{\pm}\to K^{\pm}\eta_c\eta$, & & \\
\hspace{2.2cm}$\eta\to\gamma\gamma$ & $3.05\pm 0.01$ & $-14\pm 26$ \\
\hspace{2.2cm}$\eta\to\pi^+\pi^-\pi^0$ &  $ 0.69\pm 0.01$ & $-1.8\pm 3.4$ \\
\hline
\end{tabular}
\caption{Fit results for $B$ decays independent of intermediate resonances.}
\label{tab:21}
\end{center}
\end{table}

\section{$X_1(3872)$, $X(3730)$ and $X(4014)$}

\begin{figure}[htb]
\centering
\includegraphics[height=1.8in]{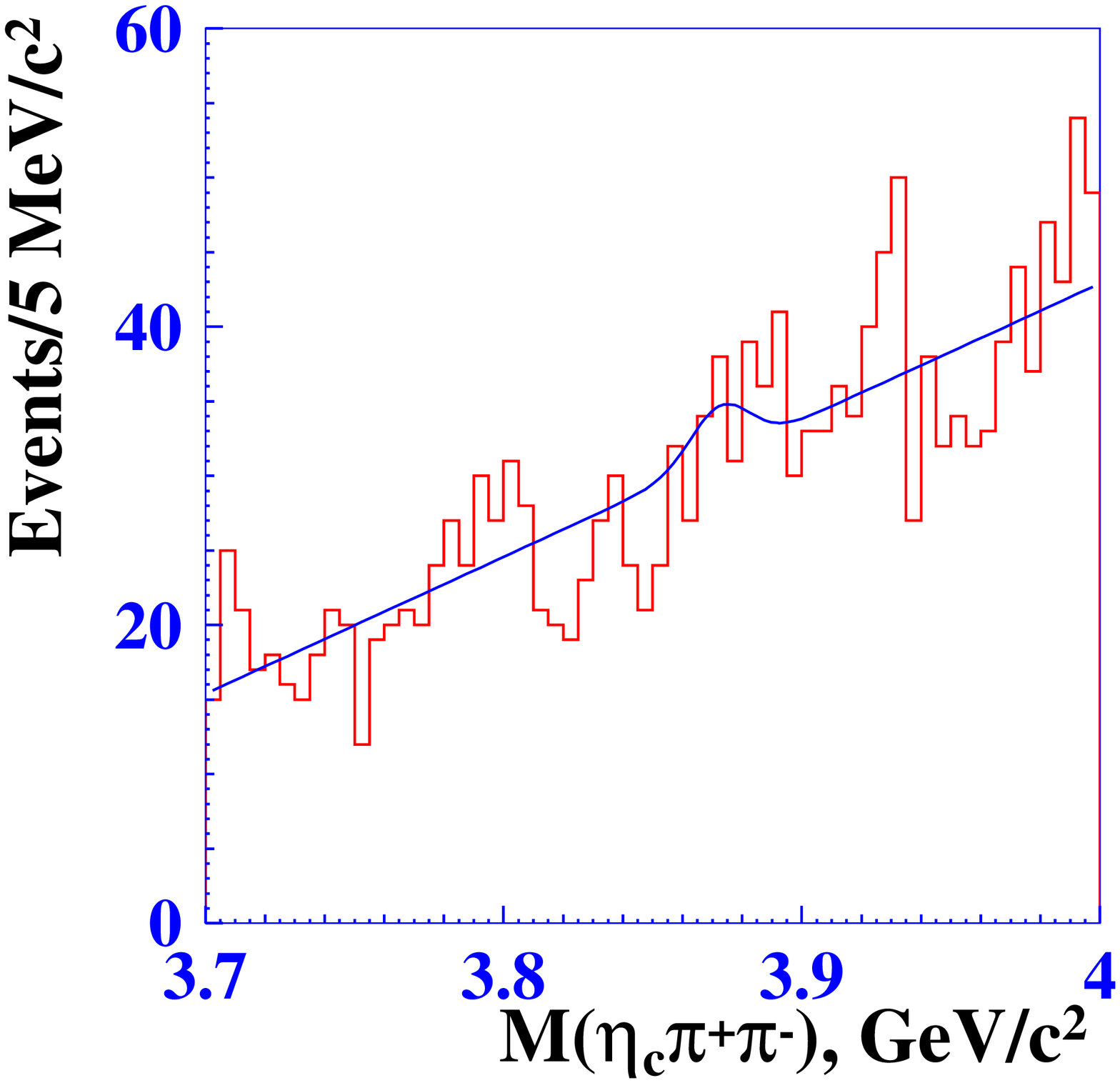}
\includegraphics[height=1.8in,origin=c,angle=0]{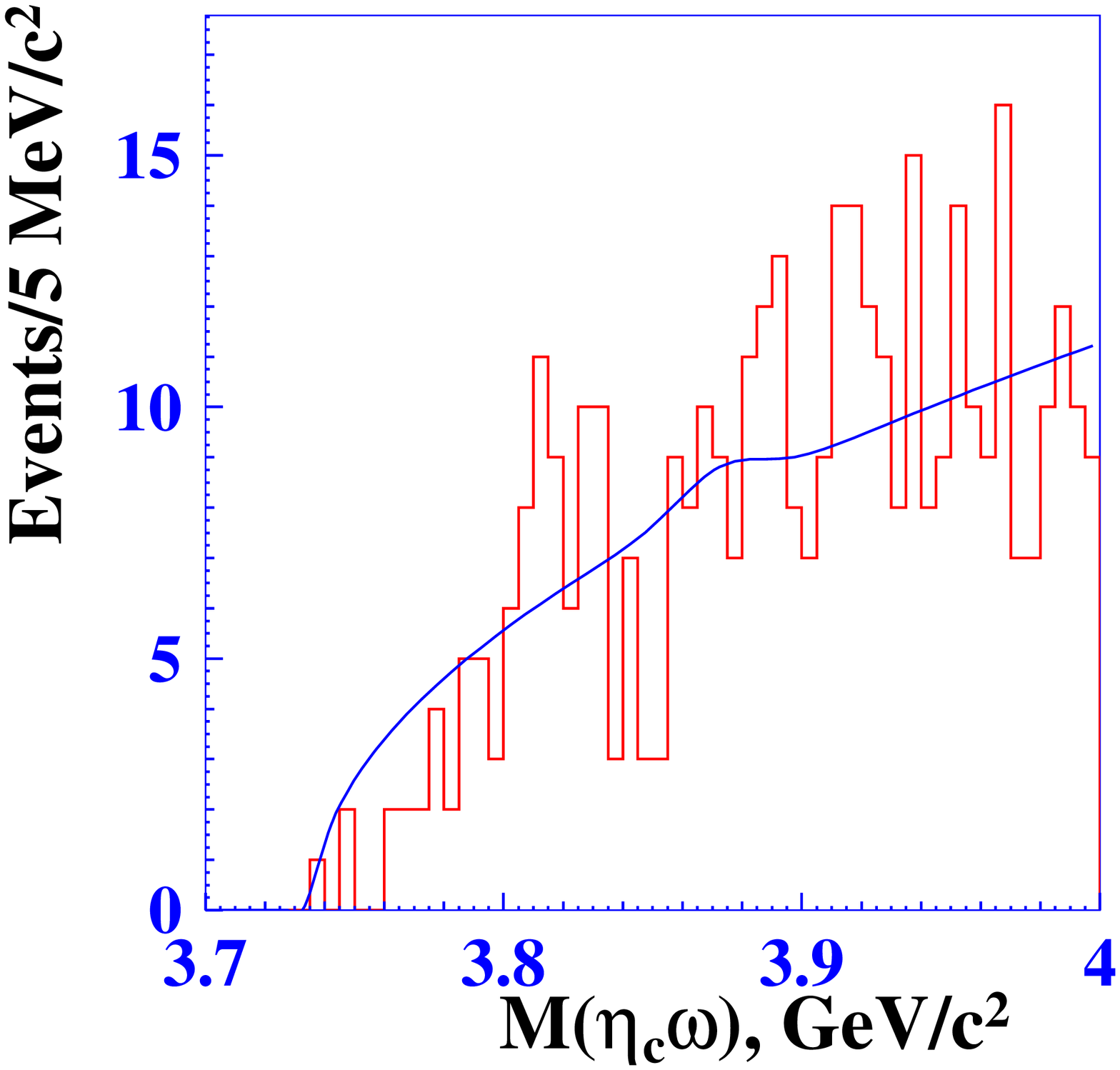}
\caption{The distributions of $\eta_c\pi^+\pi^-$ (left) and $\eta_c\omega$ (right) invariant masses in the search for the $X_1(3872)$.}
\label{pic:2}
\end{figure}

\begin{figure}[htb]
\centering
\includegraphics[height=1.8in]{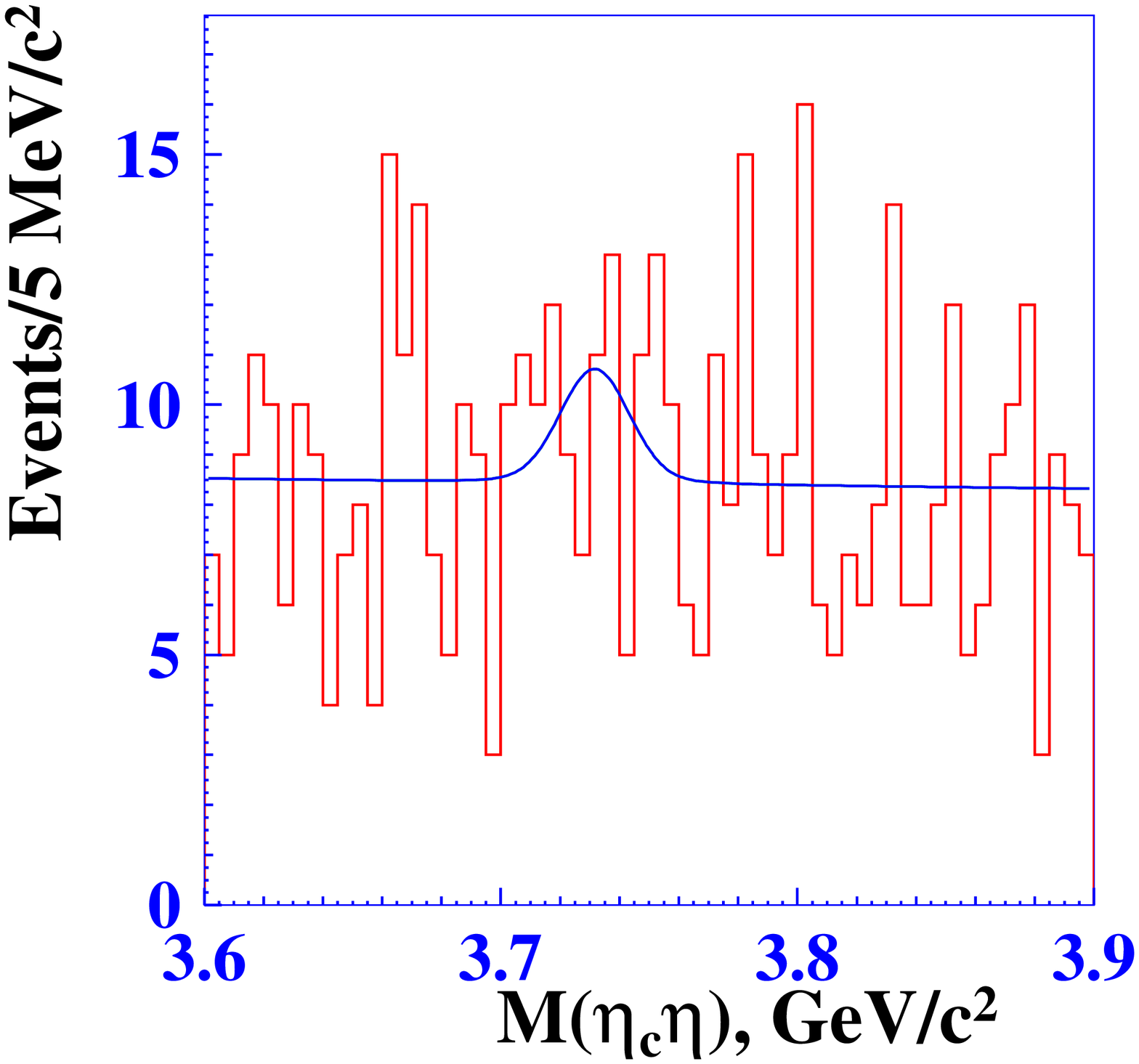}
\includegraphics[height=1.8in,origin=c,angle=0]{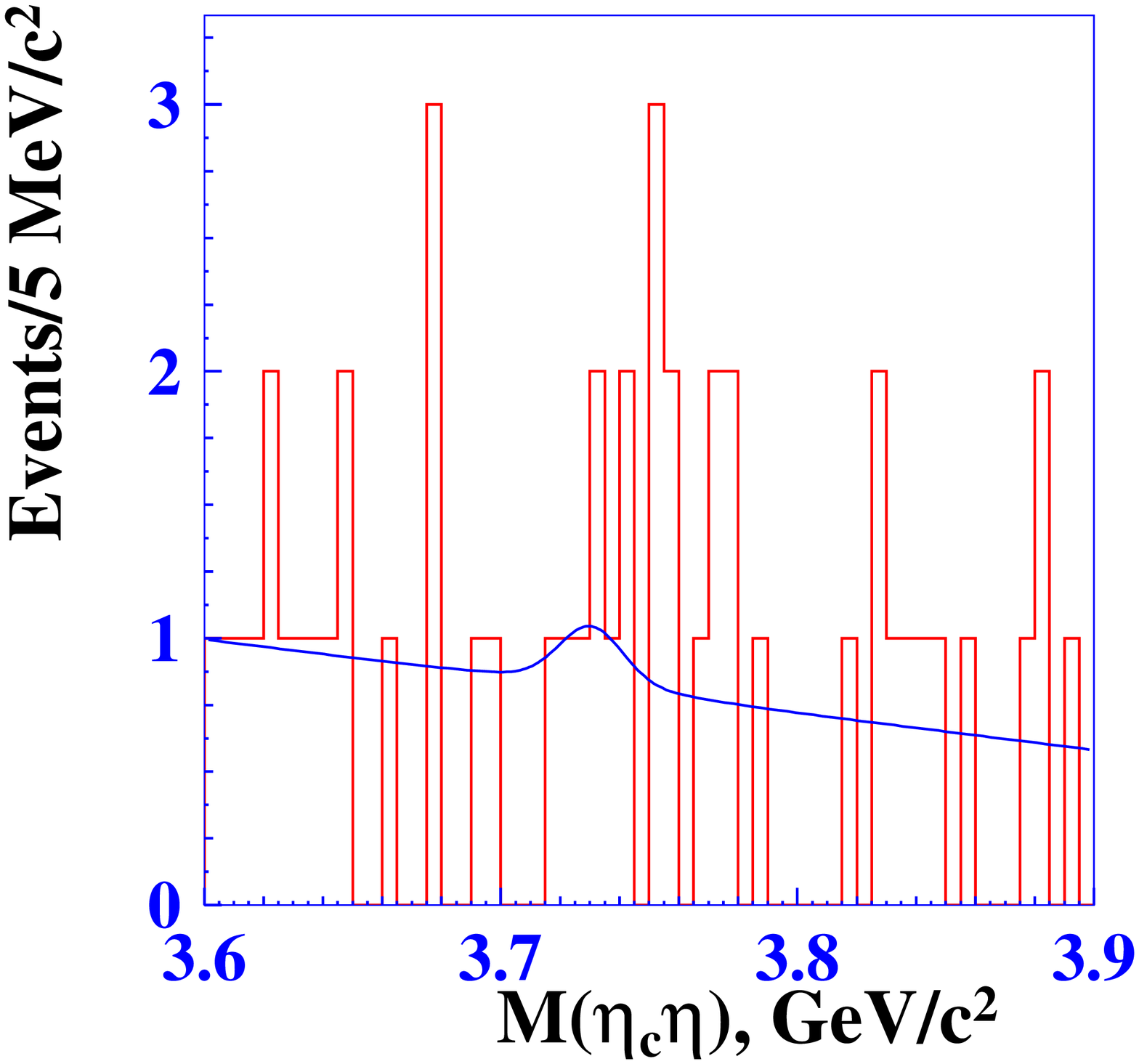}
\vfill
\includegraphics[height=1.8in]{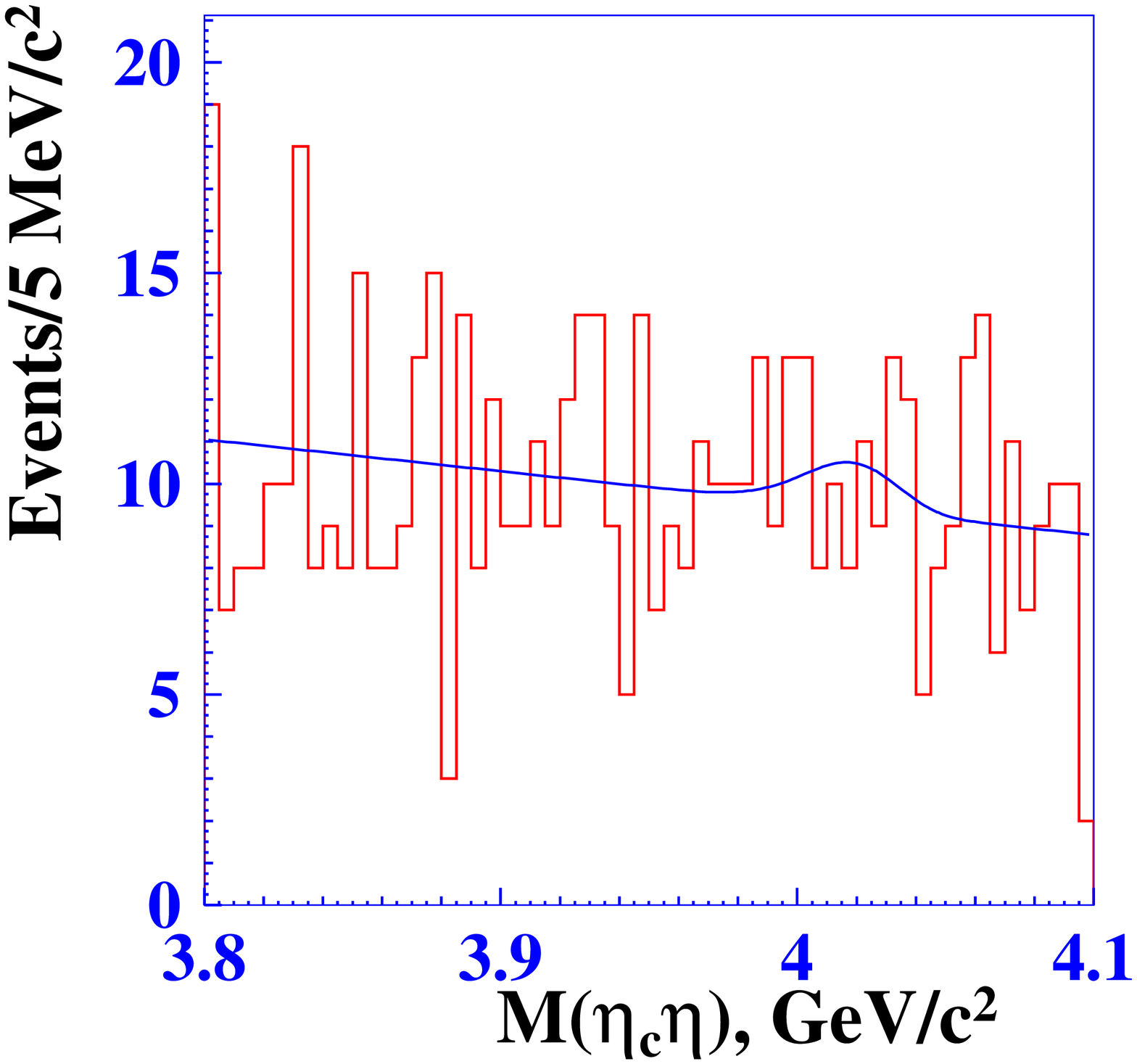}
\includegraphics[height=1.8in,origin=c,angle=0]{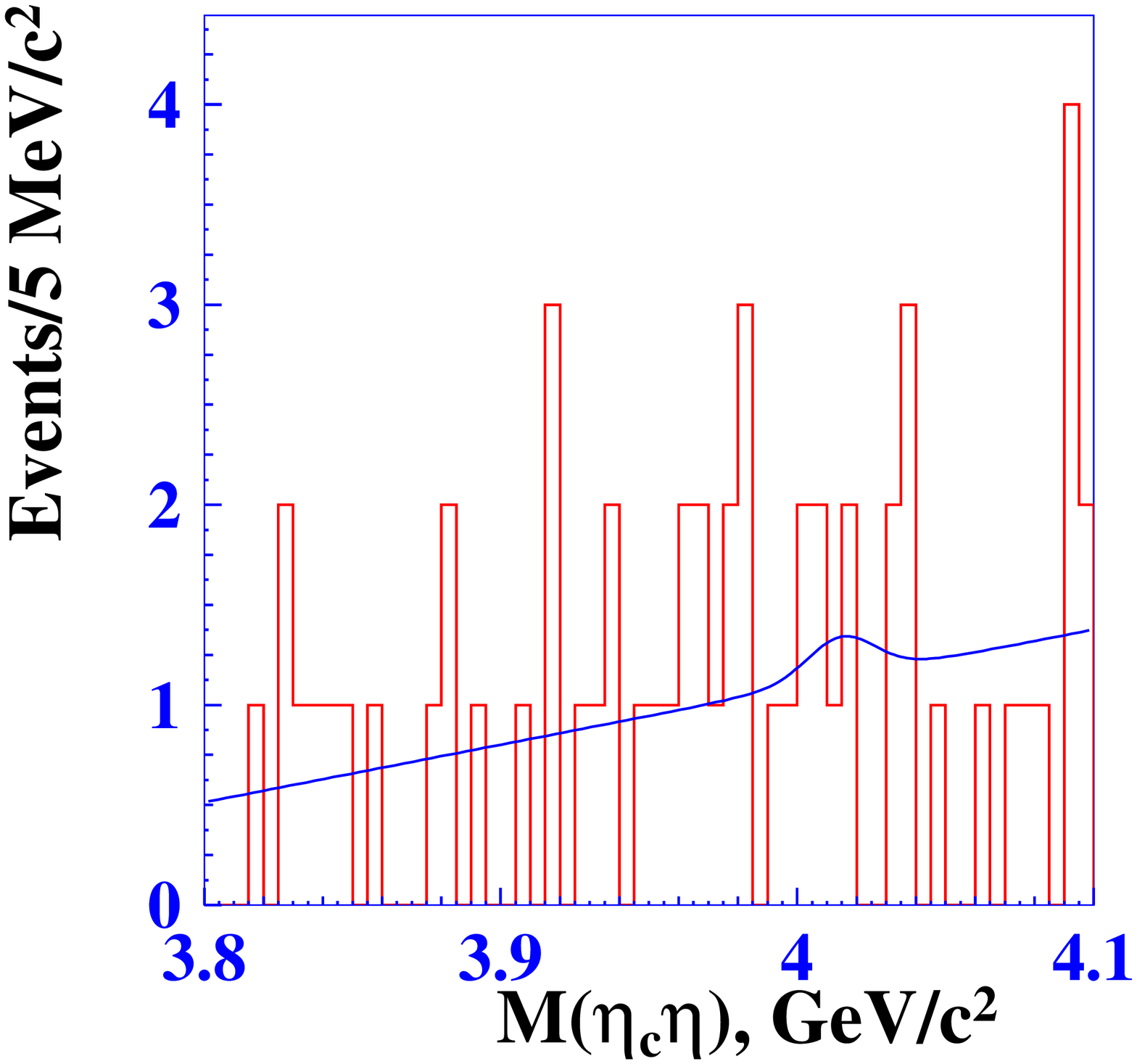}
\caption{The combined fit projections of the $\eta_c\eta$ invariant mass distributions in case of the $\eta\to\gamma\gamma$ (left) and $\eta\to\pi^+\pi^-\pi^0$ (right) modes corresponding to the search for the $X(3730)$ (top) and $X(4014)$ (bottom) resonances.}
\label{pic:3}
\end{figure}

\begin{figure}[htb]
\centering
\includegraphics[height=1.8in]{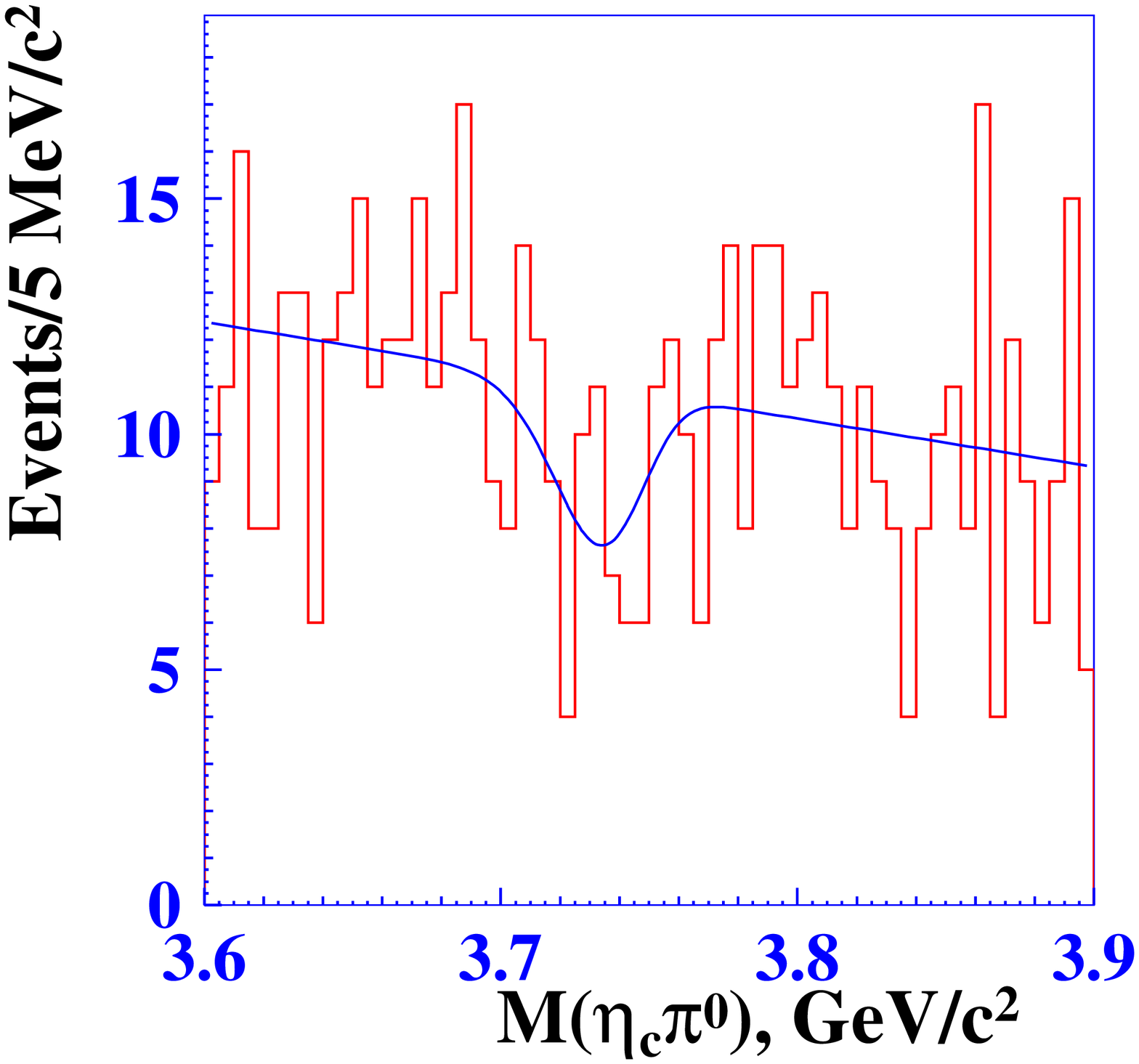}
\includegraphics[height=1.8in,origin=c,angle=0]{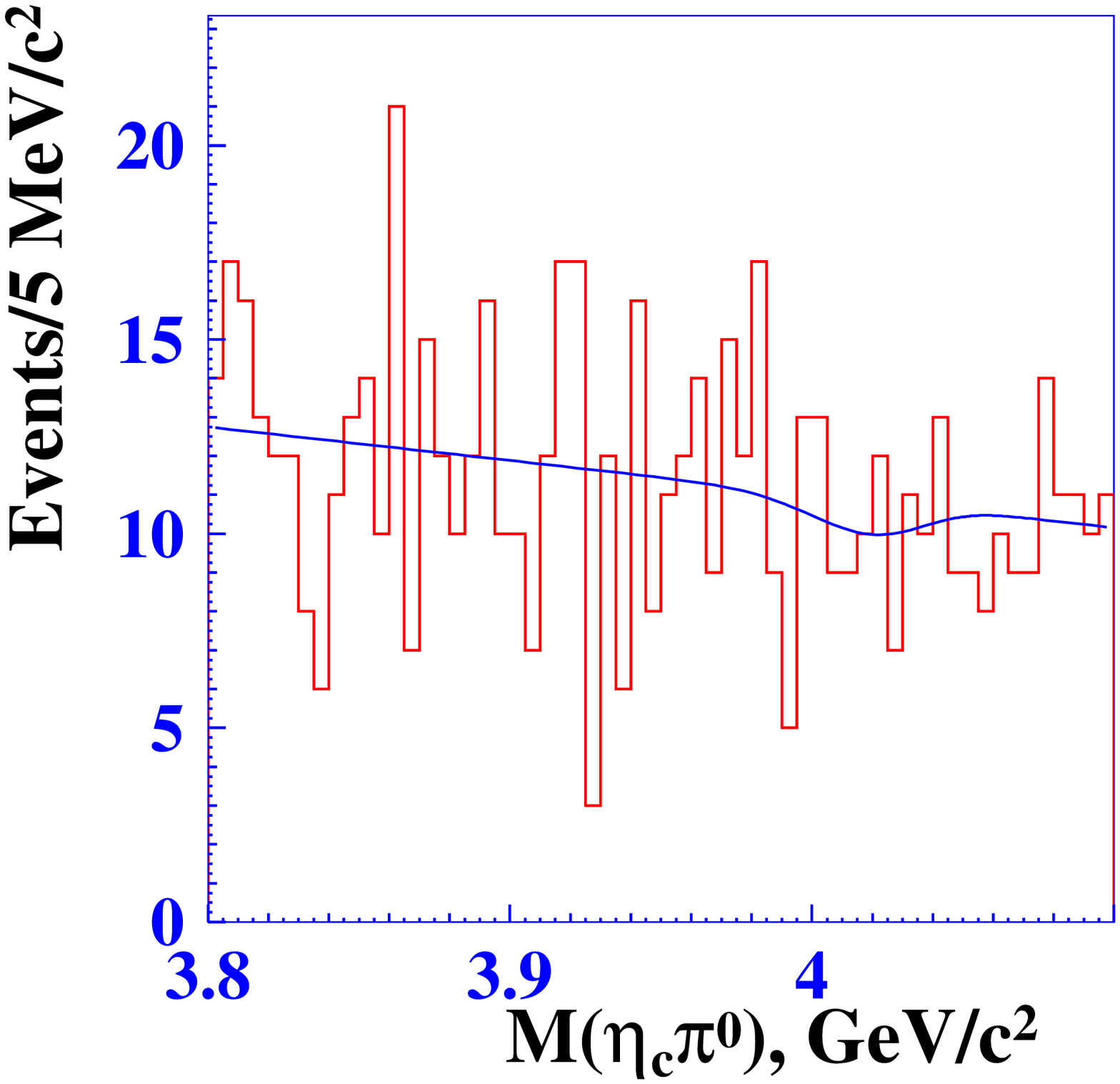}
\includegraphics[height=1.8in]{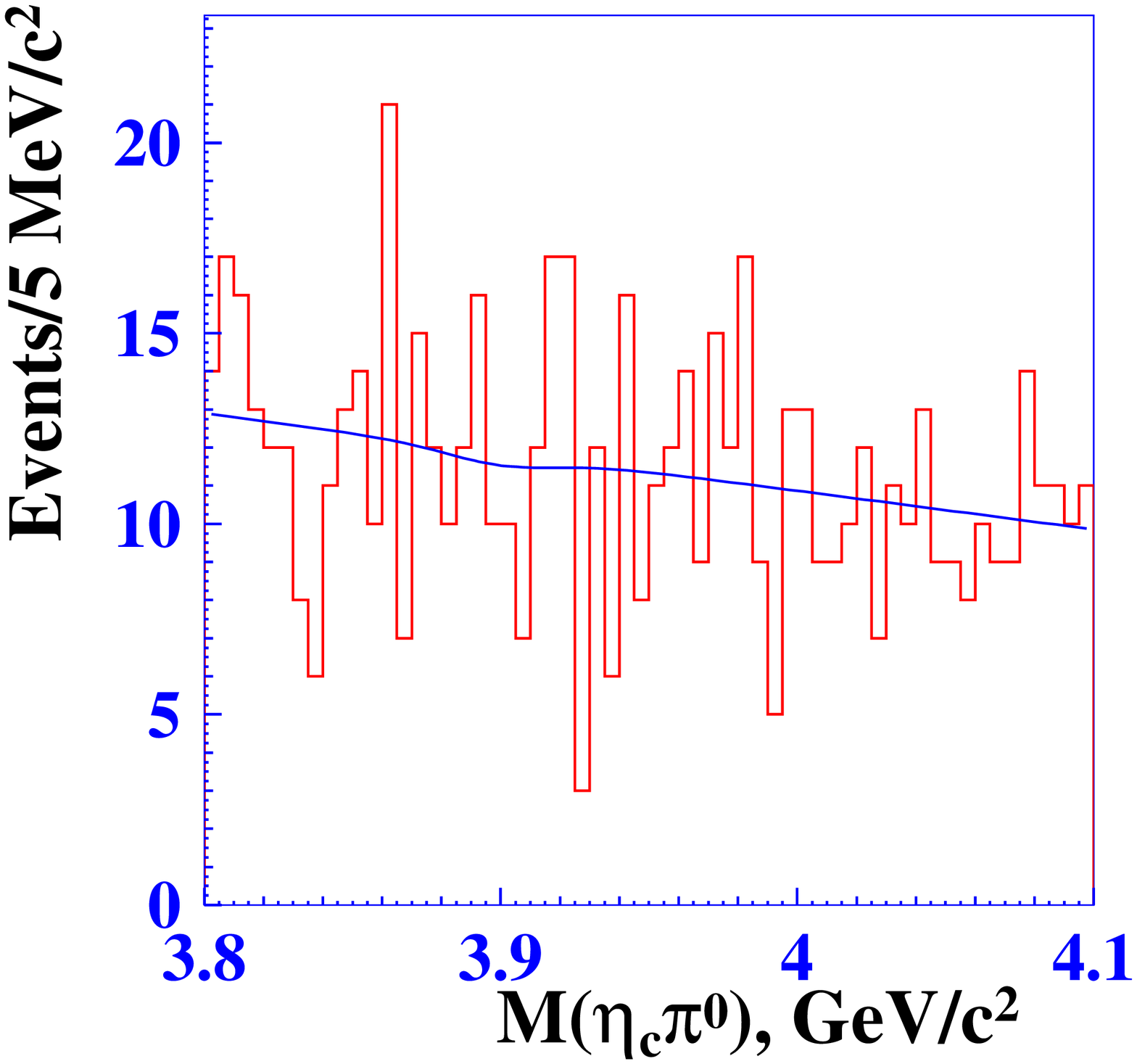}
\caption{The $\eta_c\pi^0$ invariant mass distributions corresponding to the search for the $X(3730)$ (left), $X(4014)$ (middle) and $X(3915)$ (right) resonances.}
\label{pic:4}
\end{figure}

The invariant mass distributions for ($\pi^+\pi^-$), ($\omega$), ($\eta$), and ($\pi^0$) modes are shown in Figures~\ref{pic:2}--\ref{pic:4}. The corresponding signal yields are presented in Table~\ref{tab:11}. We validate our fit procedure by applying it to the decay $B^{\pm}\to K^{\pm}\psi(2S)$, $\psi(2S)\to J/\psi\pi^+\pi^-$. This decay is similar to the ($\pi^+\pi^-$) decay except that we reconstruct the $\psi(2S)$ meson in place of the $X_1(3872)$ and the $J/\psi$ in place of the $\eta_c$. The $J/\psi$ meson, like the $\eta_c$, is reconstructed via the $K^0_SK^{\pm}\pi^{\mp}$ final state. We fit the $J/\psi\pi^+\pi^-$ invariant mass distribution and obtain the number of signal events $N_s=20.2\pm 6.5$, which corresponds to a significance of 3.5 standard deviations ($\sigma$). The expected number of events estimated using the world averages of the known branching fractions~\cite{PDG} is $22\pm 4$, which is consistent with $N_s$. 
\begin{table}[htb]
\begin{center}
\begin{tabular}{l|cc} 
Decay mode & Efficiency, \% & Yield \\
\hline
$X_1(3872)\to \eta_c\pi^+\pi^-$ & $7.95\pm 0.02$ &
$17.9\pm 16.5$ \\
$X_1(3872)\to\eta_c\omega$ & $1.92\pm 0.02$ & $6.0\pm 12.5$\\
$X(3730)\to\eta_c\eta$, & & \\
\hspace{2.5cm}$\eta\to\gamma\gamma$ & $6.57\pm 0.02$ & $13.8\pm 9.9$ \\
\hspace{2.5cm}$\eta\to\pi^+\pi^-\pi^0$ & $ 1.18\pm 0.01$ & $1.4\pm 1.0$ \\
$X(3730)\to\eta_c\pi^0$ & $6.52\pm 0.02$ & $-25.6\pm 10.4$ \\
$X(4014)\to\eta_c\eta$, & & \\
\hspace{2.5cm}$\eta\to\gamma\gamma$ & $7.09\pm 0.02$ & $8.9\pm 11.0$ \\
\hspace{2.5cm}$\eta\to\pi^+\pi^-\pi^0$ & $1.78\pm 0.01$ & $1.3\pm 1.6$ \\
$X(4014)\to\eta_c\pi^0$ & $7.55\pm 0.02$ & $-8.1\pm 13.2$ \\
$X(3915)\to\eta_c\eta$, & & \\
\hspace{2.5cm}$\eta\to\gamma\gamma$ & $6.60\pm 0.02$ & $-7.4\pm 14.5$ \\
\hspace{2.5cm}$\eta\to\pi^+\pi^-\pi^0$ & $ 1.64\pm 0.01$ & $-1.1\pm 2.1$ \\
$X(3915)\to\eta_c\pi^0$ & $6.88\pm 0.02$ & $-4.3\pm 18.1$ \\
\hline
\end{tabular}
\caption{Fit results for the $X_1(3872)$, $X(3730)$, $X(4014)$ and $X(3915)$ resonances.}
\label{tab:11}
\end{center}
\end{table}

\section{$Z(3900)^0$ and $Z(4020)^0$}

\begin{figure}[htb]
\centering
\includegraphics[height=1.8in]{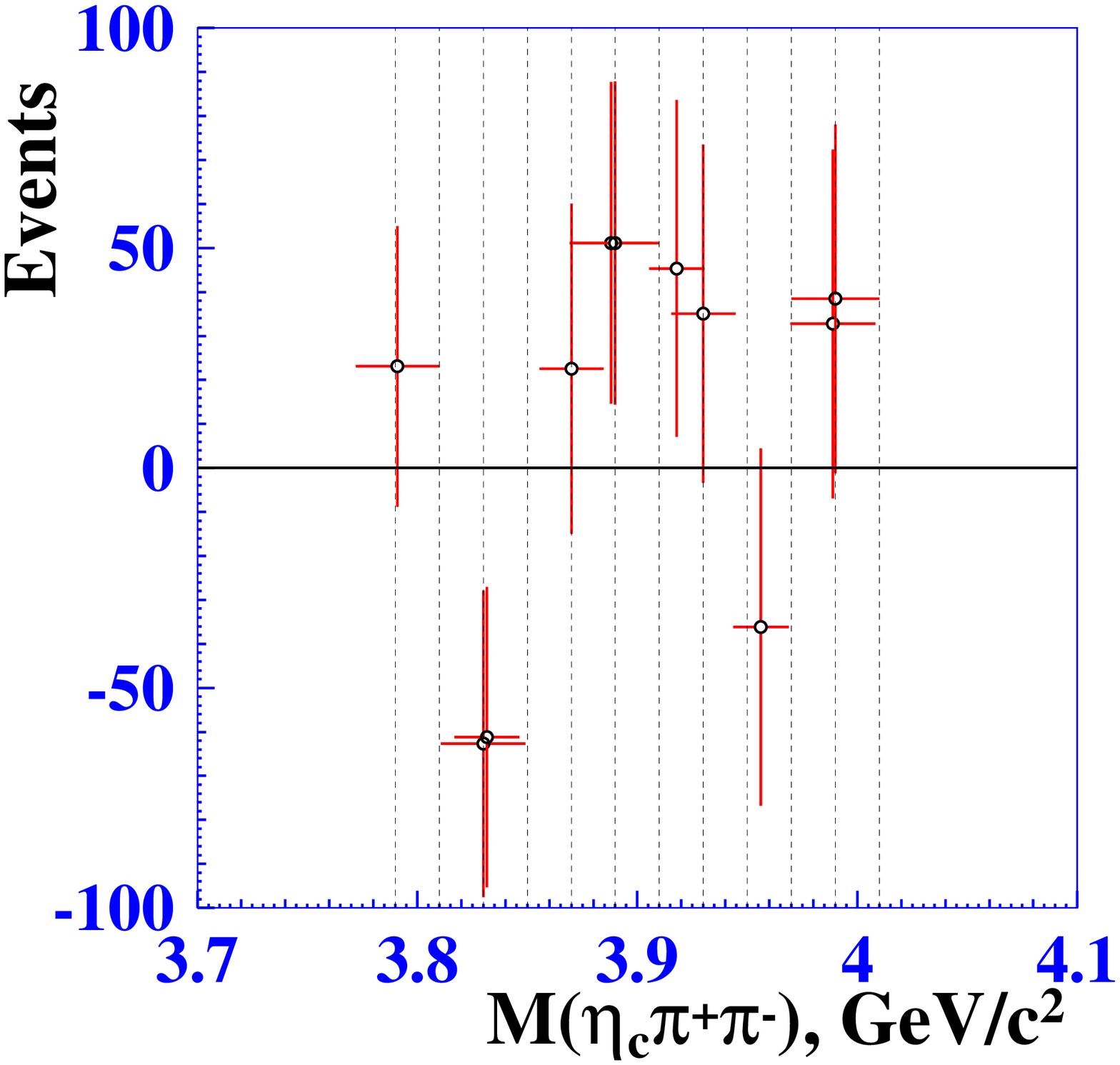}
\includegraphics[height=1.8in,origin=c,angle=0]{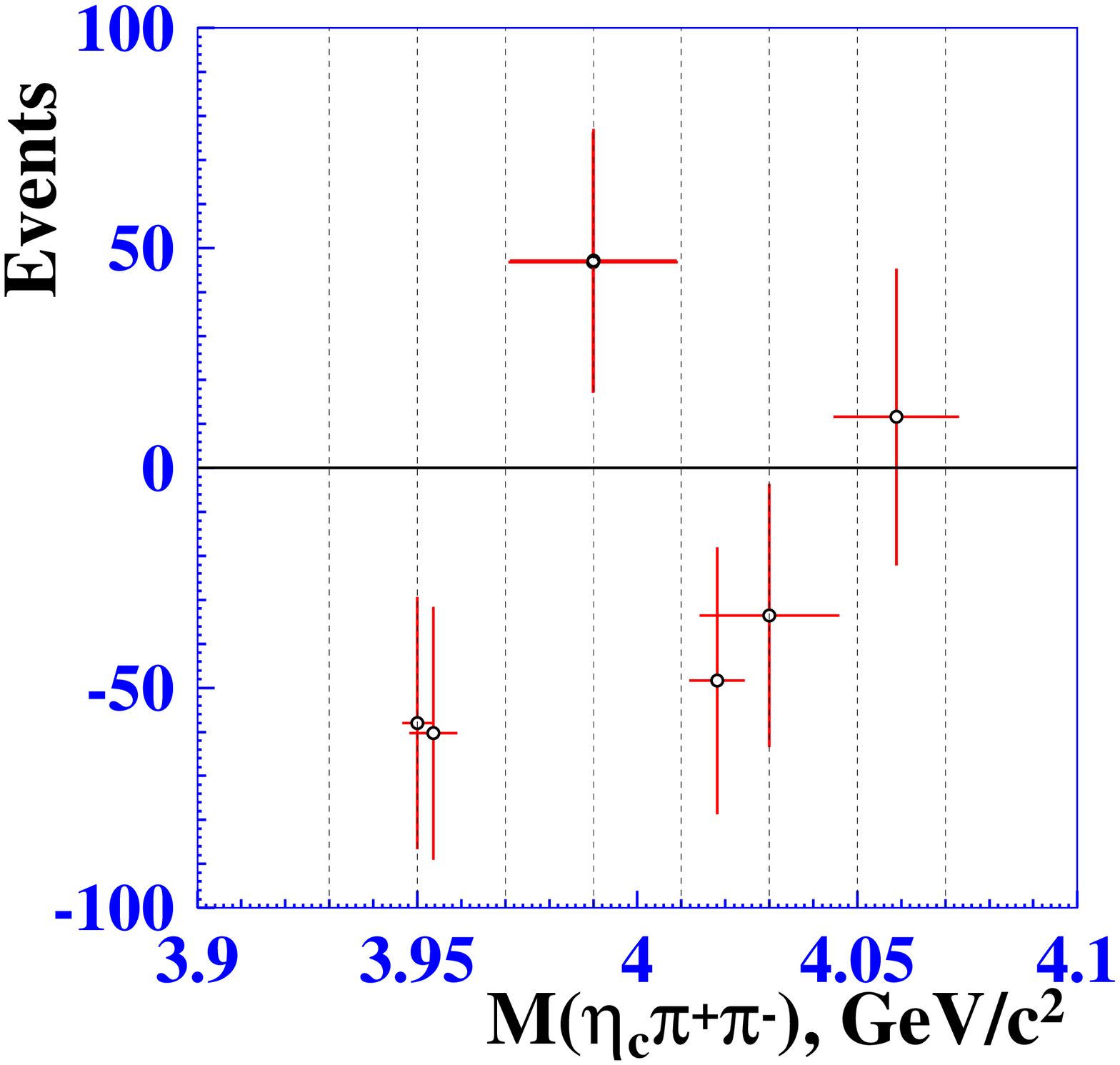}
\caption{Dependence of the signal yield of the $Z(3900)^0$ (left) and $Z(4020)^0$ (right) on the mass bin. The mass bin is a $20$ MeV/$c^2$ window to which the mass is confined and scanned in $20$ MeV/$c^2$ steps across the fit range.}
\label{pic:33}
\end{figure}

We perform a sequence of binned maximum likelihood fits of the $\eta_c\pi^+\pi^-$ invariant mass using the convolution of a Breit-Wigner and a Gaussian for the signal and a linear polynomial for the background.
The Gaussian models the detector resolution, which is assumed to be similar to that obtained in Ref.~\cite{Vinok} and equal to $9.8$ MeV/$c^2$. The Breit-Wigner mass is confined to a $20$ MeV/$c^2$ window (the so-called mass bin) that is scanned in $20$ MeV/$c^2$ steps across the range $(3.79-4.01)$ GeV/$c^2$ for the $Z(3900)^0$ and $(3.93-4.07)$ GeV/$c^2$ for the $Z(4020)^0$. The width is fixed to the weighted mean of the previously measured values ($35$ MeV/$c^2$ for the $Z(3900)^0$ and $12$ MeV/$c^2$ for the $Z(4020)^0$). The obtained signal yield is shown in Figure~\ref{pic:33}. 

\section{$X(3915)$}

\begin{figure}[htb]
\centering
\includegraphics[height=1.8in]{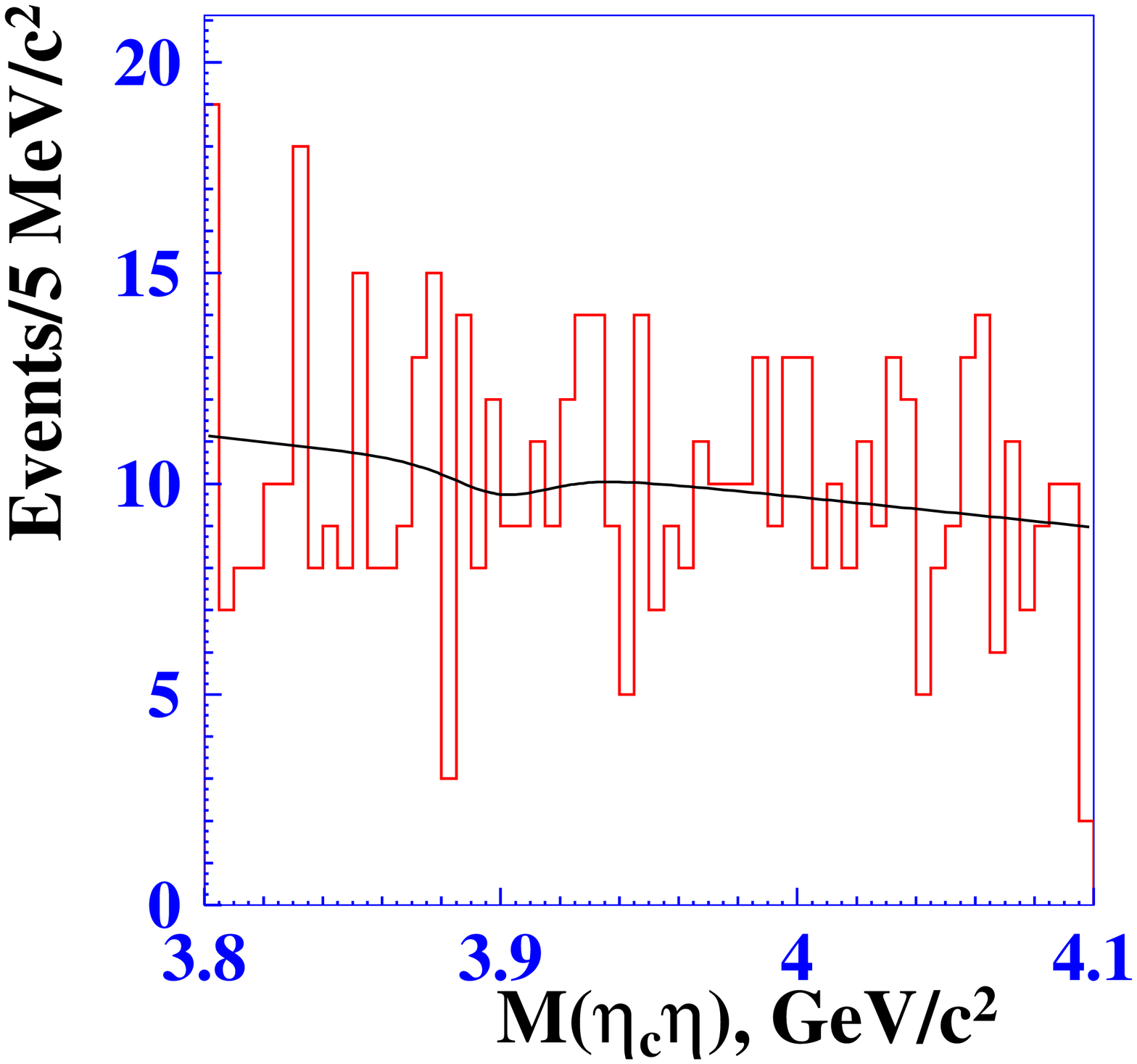}
\includegraphics[height=1.8in,origin=c,angle=0]{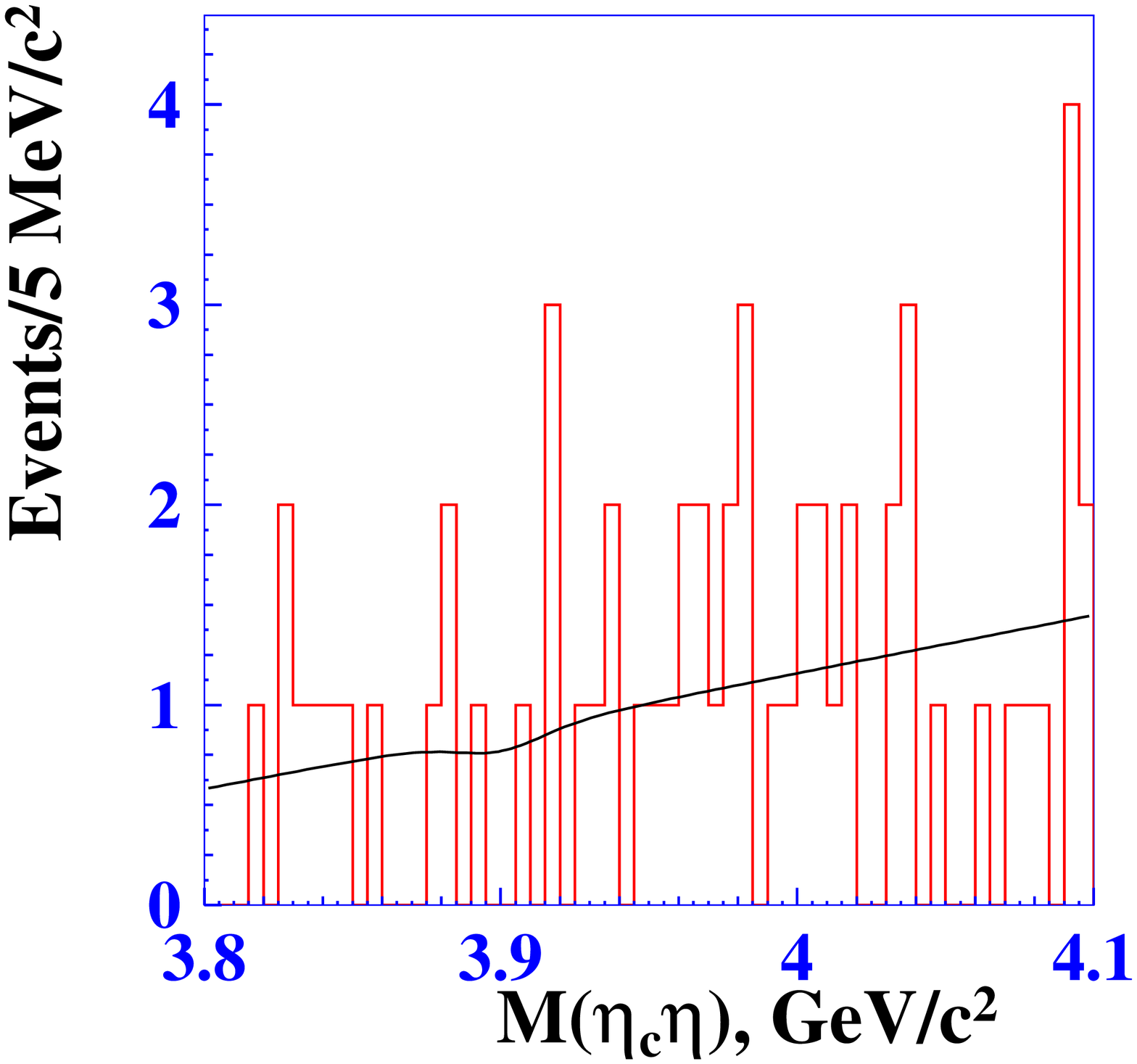}
\caption{The combined fit projections of the $\eta_c\eta$ invariant mass distributions in case of the $\eta\to\gamma\gamma$ (left) and $\eta\to\pi^+\pi^-\pi^0$ (right) modes corresponding to the search for the $X(3915)$ resonance.}
\label{pic:39151}
\end{figure}

The $\eta_c\eta$ and $\eta_c\pi^0$ invariant mass distributions are shown in Figures~\ref{pic:39151} and~\ref{pic:4}. The fit results are summarized in Table~\ref{tab:11}.

\clearpage
\section{Results}

Our data sample does not permit us to measure the branching products of production and decay of the states listed above nor the $B$ decay branching fractions so we set upper limits instead, taking into account the statistical and systematic uncertainties. Upper limits on the branching fractions and products for all the studied decay modes are shown in Tables~\ref{tab:6} and~\ref{tab:5}.
\begin{table}[htb]
\begin{center}
\begin{tabular}{l|c} 
Decay mode & Upper limit (90\% C.L.)\\
\hline
$B^{\pm}\to K^{\pm}\eta_c\pi^+\pi^-$ & $3.9\times 10^{-4}$\\
$B^{\pm}\to K^{\pm}\eta_c\omega$ & $5.3\times 10^{-4}$\\
$B^{\pm}\to K^{\pm}\eta_c\eta$, & $2.2\times 10^{-4}$\\
$B^{\pm}\to K^{\pm}\eta_c\pi^0$ & $6.2\times 10^{-5}$\\ 
\hline
\end{tabular}
\caption{Results of branching fraction measurements for the $B$ decays without an intermediate resonance.}
\label{tab:6}
\end{center}
\end{table}
\begin{table}[htb]
\begin{center}
\begin{tabular}{ll|c} 
Resonance & Decay mode & Upper limit (90\% C.L.)\\
\hline
$X_1(3872)$ & $\eta_c\pi^+\pi^-$ & $3.0\times 10^{-5}$\\
 & $\eta_c\omega$ & $6.9\times 10^{-5}$\\
\hline
$X(3730)$ & $\eta_c\eta$ & $4.6\times 10^{-5}$\\
 & $\eta_c\pi^0$ & $5.7\times 10^{-6}$\\ 
\hline
 $X(4014)$ & $\eta_c\eta$ & $3.9\times 10^{-5}$\\
 & $\eta_c\pi^0$ & $1.2\times 10^{-5}$\\ 
\hline
$Z(3900)^0$ & $\eta_c\pi^+\pi^-$ & $4.7\times 10^{-5}$\\
$Z(4020)^0$ &  & $1.6\times 10^{-5}$\\
\hline
 $X(3915)$ & $\eta_c\eta$ & $3.3\times 10^{-5}$\\
 & $\eta_c\pi^0$ & $1.8\times 10^{-5}$\\ 
\hline
\end{tabular}
\caption{\label{tab:5} Results of branching fraction measurements for the $B$ decays containing an intermediate exotic resonance. For $Z(3900)^0$ and $Z(4020)^0$ resonances the results are shown under the assumption that the masses are close to those of their charged partners.}
\label{tab:5}
\end{center}
\end{table}

In case of the $Z(3900)^0$ and $Z(4020)^0$ mass scan, no significant signal is seen in any of the invariant mass bins. For the $Z(3900)^0$ resonance, we set upper limits in the range $(1.8-4.7)\times 10^{-5}$ for the mass region $(3.79-4.01)$ GeV/$c^2$. For the $Z(4020)^0$ resonance, we set upper limits in the range $(1.6-3.7)\times 10^{-5}$ for the mass region $(3.93-4.07)$ GeV/$c^2$. If we assume that the $Z(3900)^0$ and $Z(4020)^0$ masses are close to those of their charged partners, we obtain the upper limits on the product branching fractions shown in Table~\ref{tab:5}.

The results are published in~\cite{Vinok2}.

%\clearpage


\begin{thebibliography}{99}

\bibitem{Belle1} S.~K.~Choi {\it et al.}  [Belle Collaboration], \emph{Phys.\ Rev.\ Lett.}  {\bf 91} (2003) 262001.
\bibitem{Molecule} M.~Suzuki, \emph{Phys.\ Rev.} D {\bf 72} (2005) 114013.
\bibitem{LHCb2} R.~Aaij {\it et al.}  [LHCb Collaboration], \emph{Phys.\ Rev.\ Lett.}  {\bf 110} (2013) 222001.
\bibitem{zbelle} Z.~Q.~Liu {\it et al.}  [Belle Collaboration], \emph{Phys.\ Rev.\ Lett.}  {\bf 110} (2013) 252002.
\bibitem{zbes1} M.~Ablikim {\it et al.}  [BESIII Collaboration], \emph{Phys.\ Rev.\ Lett.}  {\bf 110} (2013) 252001.
\bibitem{zcleo} T.~Xiao, S.~Dobbs, A.~Tomaradze and K.~K.~Seth, \emph{Phys.\ Lett.} B {\bf 727} (2013) 366.
\bibitem{zbes3} M.~Ablikim {\it et al.}  [BESIII Collaboration], \emph{Phys.\ Rev.\ Lett.}  {\bf 111} (2013) 24,  242001.
\bibitem{zbes4} M.~Ablikim {\it et al.}  [BESIII Collaboration], \emph{Phys.\ Rev.\ Lett.}  {\bf 112} (2014) 13,  132001.
\bibitem{x39151} K.~Abe {\it et al.}  [Belle Collaboration], \emph{Phys.\ Rev.\ Lett.}  {\bf 94} (2005) 182002.
\bibitem{x39152} B.~Aubert {\it et al.}  [BaBar Collaboration], \emph{Phys.\ Rev.\ Lett.}  {\bf 101} (2008) 082001.
\bibitem{x39156} T.~Branz, T.~Gutsche and V.~E.~Lyubovitskij, \emph{Phys.\ Rev.} D {\bf 80} (2009) 054019; X.~Liu, Z.~G.~Luo, Y.~R.~Liu and S.~L.~Zhu, \emph{Eur.\ Phys.\ J.} C {\bf 61} (2009) 411; W.~H.~Liang, R.~Molina and E.~Oset, \emph{Eur.\ Phys.\ J.} A {\bf 44} (2010) 479; X.~Liu, Z.~G.~Luo and Z.~F.~Sun, \emph{Phys.\ Rev.\ Lett.}  {\bf 104} (2010) 122001; S.~L.~Olsen, \emph{Phys.\ Rev.\ D} {\bf 91} (2015) 5,  057501.
\bibitem{PDG} K.~A.~Olive {\it et al.}  [Particle Data Group Collaboration], \emph{Chin.\ Phys.} C {\bf 38} (2014) 090001.
\bibitem{Vinok} A.~Vinokurova {\it et al.}  [Belle Collaboration], \emph{Phys.\ Lett.} B {\bf 706} (2011) 139.
\bibitem{Vinok2} A.~Vinokurova {\it et al.}  [Belle Collaboration], \emph{JHEP} {\bf 06} (2015) 132.
\end{thebibliography}
\end{document}